\documentclass[preprint,showpacs,preprintnumbers,amsmath,amssymb]{revtex4}
\usepackage{epsfig}
\usepackage{amssymb}

\begin{document}
\title{Nematic-Isotropic Spinodal Decomposition
Kinetics of Rod-like Viruses. }
\author{M. Paul Lettinga\dag, Kyongok Kang\dag, Peter Holmqvist\dag, Arnout Imhof\ddag, Didi Derks\ddag and Jan K. G. Dhont\dag}
\address{\dag
Forschungs Zentrum J\"{u}lich, IFF, Weiche Materie, J\"{u}lich,
D-52425 J\"{u}lich, Germany }
\address{\ddag
Soft Condensed Matter, Debye Institute, Utrecht University,
Princetonplein 5, 3584 CC Utrecht, The Netherlands}


%
%

\bibliographystyle{unsrt}
\begin{abstract}
We investigate spinodal decomposition kinetics of an initially
nematic dispersion of rod-like viruses (fd virus). Quench
experiments are performed from a flow-stabilized homogeneous
nematic state at high shear rate into the two-phase
isotropic-nematic coexistence region at zero shear rate. We
present experimental evidence that spinodal decomposition is
driven by orientational diffusion, in accordance with a very
recent theory.
\end{abstract}

\maketitle

\newcommand{\hg}{\mbox{$h(\mathbf{r})$}}
\newcommand{\rr}{\mbox{$\mathbf{r}$}}
\newcommand{\kk}{\mbox{$\mathbf{k}$}}
\newcommand{\hkk}{\mbox{$\hat{\mathbf{k}}$}}
\newcommand{\qq}{\mbox{$\mathbf{q}$}}
\newcommand{\uu}{\mbox{$\hat{\mathbf{u}}$}}
\newcommand{\nn}{\mbox{$\hat{\mathbf{n}}$}}
\newcommand{\rot}{\mbox{$\hat{\mathcal{R}}$}}
\newcommand{\dg}{\mbox{$\dot{\gamma}$}}

\section{Introduction}

Systems that are quenched into a state where at least one order
parameter is unstable undergo spinodal phase separation. Here, the
initially homogeneous system is unstable against fluctuations of
arbitrary small amplitude, and phase separation sets in
immediately after a quench. In the initial stage of phase
separation an interconnected "labyrinth structure" of regions with
somewhat higher and lower values of the order parameter is
observed. For systems containing spherical particles the relevant
order parameter is the concentration. As Onsager showed in
1949\cite{Onsager49}, the situation is different when the
particles are not spherical in shape, i.e. disk-like or elongated
particles. Here the system can become unstable or meta-stable with
respect to fluctuations in \emph{orientation}. These orientational
fluctuations drive concentrations differences, resulting in a
phase with high concentration and orientational order, the
\emph{nematic} phase, and a phase with low concentration and no
orientational order, the \emph{isotropic} phase. For very long and
thin rods with short-ranged repulsive interactions, the binodal
concentrations, i.e. the concentrations of the isotropic and
nematic phases in equilibrium after phase separation is completed,
have been determined using different approximations in minimizing
Onsager's functional for the free energy (see ref. \cite{Vroege92}
and references therein), while for shorter rods computer
simulations have been performed to obtain binodal concentrations
\cite{Bolhuis97,Graf99}. The spinodal concentration, where the
isotropic phase becomes unstable has been obtained
\cite{Onsager49,Kayser78}.

Recently, a microscopic theory was developed by one of the
authors,  describing the initial stage of the \emph{kinetics} of
spinodal decomposition (SD) \cite{Dhont05}. It is shown there that
demixing is dominated by rotational diffusion and not by
translational diffusion as suggested in earlier work
\cite{winters00,maeda89}. This is in line with results based on
Ginzburg-Landau equations of motion, where the importance of the
coupling between concentration and orientation was studied
\cite{Liu96,Matsuyama00} for rod-polymer mixtures.

Signatures of SD have been observed for suspensions of boehmite
rods, by homogenizing a phase separated system and sequential
polarization microscopy and Small Angle Light Scattering (SALS)
measurements \cite{vanbruggen99a}. For such experiments, however,
there is always an experimental lapse time between homogenization
and the first moment of observation. In these experiments the
initial state is not well defined. Ideally one would like to
perform a concentration quench from low or high concentration into
the two-phase region, where the initial state is isotropic or
nematic, respectively. In a recent paper such a kind of 'quench'
was performed by inducing polymerization of short actin chains
\cite{Viamontes05}. Alternatively, external fields like shear flow
\cite{Lenstra01} or magnetic fields \cite{Tang93} can be used to
achieve well-defined quenches. Switching on or turning off such an
external field can take the initially homogeneous system into
either an unstable or meta-stable state.

In this paper we induce a nematic phase with a well defined
director by imposing shear flow to a dispersion of colloidal rods.
At a sufficiently high shear rate, the fully nematic, homogeneous
state is stable. The shear flow is then suddenly switched off,
after which the system becomes either unstable or meta-stable. As
a system we use suspensions of \emph{fd}-viruses, which are
mono-disperse and very long and thin somewhat flexible particles.
The equilibrium phase behavior for these semi-flexible rods as far
as the binodal points are concerned, is well understood on the
basis of Onsager theory, extended to include charge and
flexibility \cite{Tang95,Chen93}. Polymer is added to the
dispersion in order to widen the region of isotropic-nematic phase
coexistence, which renders phase separation experiments feasible
\cite{Dogic04a}. In a previous study we obtained the spinodal
point that separates the unstable and meta-stable region relevant
for the initial state in the experiments described in the present
paper \cite{Lettinga05c}. In the present paper, experiments are
performed at concentrations such that the cessation of shear will
render the system unstable. We interpret our data on the basis of
the recent microscopic theory \cite{Dhont05} mentioned above.
Experimental evidence is given that phase separation is indeed
driven by orientational ordering, which enslaves the
concentration. We use confocal microscopy to confirm that demixing
is indeed proceeding via spinodal decomposition, rather than
nucleation and growth, through the observation of an initial
interconnected structure of inhomogeneities. In addition we also
use SALS experiments because these have a better time resolution
and a better statistics. The present experiments are qualitative
in the sense that only the specific wave vector dependence of the
unstable eigen mode is discussed, without systematically varying
the amount of added dextran which leads to attraction between the
rods.

This paper is organized as follows. First we give a brief overview
of the microscopic theory on SD of rod dispersions \cite{Dhont05},
leading to predictions that will be tested experimentally. After
the section on materials and methods, experimental results are
presented for both techniques. In the discussion we analyze our
results using the predictions of the microscopic theory.

\section{Theory}\label{theory}

The time-evolution of the probability density function (pdf) of
the orientations and positions of an assembly of $N$ rods is
described by the so-called Smoluchowski equation. From this
microscopic equation of motion, an equation of motion for the
number density $\rho(\rr,\uu,t)$ of rods at $\rr$ with orientation
$\uu$ at time $t$ can be derived by integration \cite{Dhont05},
\begin{eqnarray}\label{P1}
\frac{\partial }{\partial t}\,\rho(\rr,\uu,t)&=&
\frac{3}{4}\bar{D} \nabla \cdot \mathbf{D}(\uu)\cdot \left\{
\nabla \rho(\rr,\uu,t)
- \beta\,\rho(\rr,\uu,t)\,\bar{\mathbf{F}}(\rr,\uu,t)\right\} \nonumber \\
&&\!\!\!+\,D_{r} \rot\cdot\left\{ \rot \rho(\rr,\uu,t)
-\beta\,\rho(\rr ,\uu ,t)\,\bar{\mathbf{T}}(\rr,\uu,t) \right\}
\;.
\end{eqnarray}
Here, $\bar{D}$ and $D_{r}$ are the orientationaly averaged
translational diffusion coefficient and rotational diffusion
coefficient of a non-interacting rod, respectively. The
orientational dependence of the translational diffusion
coefficient of a non-interacting rod is described by the tensor
$\mathbf{D}(\uu)\;=\;\left[\,\hat{\mathbf{I}}+\uu\uu\,\right]$.
Furthermore,
$\rot_{j}\,(\cdots)=\uu_{j}\times\nabla_{\mbox{\small{$\hat{\mathbf{u}}_{j}$}}}\,(\cdots)$
is the rotational operator with respect to $\uu_{j}$, where
$\nabla_{\mbox{\small{$\hat{\mathbf{u}}_{j}$}}}$ is the gradient
operator with respect to the cartesian coordinates of $\uu_{j}$.
For very long and thin rods with hard-core interactions, the
average force $\bar{\mathbf{F}}$ and torque $\bar{\mathbf{T}}$ on
a rod with position $\rr$ and orientation $\uu$ due to
interactions with other rods are given by,
\begin{eqnarray}\label{P4}
\bar{\mathbf{F}}(\rr,\uu,t)\,=\,-\nabla\,V^{\mbox{eff}}(\rr,\uu,t)\;\;,\;\;\mbox{and}\;\;,\;\;
\bar{\mathbf{T}}(\rr,\uu,t)\,=\,-\rot\,V^{\mbox{eff}}(\rr,\uu,t)\;,
\end{eqnarray}
where the "effective potential",
\begin{eqnarray}
V^{\mbox{eff}}\;=\;\mbox{\small{$\frac{1}{2}$}}\,DL^{2}\beta^{-1}
\oint d\uu'\,\mid\!\uu\times\uu'\!\mid\int_{-1}^{1}
dl\int_{-1}^{1}
dl'\;\rho(\rr+\mbox{\small{$\frac{1}{2}$}}L\,l\,\uu+\mbox{\small{$\frac{1}{2}$}}L\,l'\,\uu',\uu',t)\;,
\end{eqnarray}
has been introduced earlier by Doi and Edwards \cite{Doi86}.

In order to describe initial decomposition kinetics, the density
$\rho(\rr,\uu,t)$ is written as,
\begin{eqnarray} \label{P6}
\rho(\rr,\uu,t)\;=\;\bar{\rho}\,P_{0}(\uu,t)+\delta\rho(\rr,\uu,t)\;,
\end{eqnarray}
where $\delta\rho$ is the small deviation with respect to the
initial probability density function $\bar{\rho}P_{0}(\uu,t=0)$,
with $\bar{\rho}=N/V$ the average number density of rods. Note
that $P_{0}$ is generally a function of time, which reflects the
temporal evolution of alignment of the otherwise homogeneous
system. Although we treat in this paper a quench from the nematic
state, we will now proceed by assuming that the initial state is
isotropic. Non-isotropic initial states require numerical
analysis, since an appropriate (non-linear) equation of motion for
$P_0$ should be solved simultaneously to the equation of motion
for $\delta \rho$ \cite{Dhont05}. The general features of demixing
are probably not very different for the different initial states.

During the initial stage of demixing, $\delta\rho$ can be expanded
up to second order in spherical harmonics as,
\begin{eqnarray} \label{pdf2}
\delta\rho(\rr,\uu,t)\;=\;A_{0}(\rr,t)+\mathbf{A}_{2}(\rr,t):\uu\uu\;.
\end{eqnarray}
The scalar $A_{0}$ is proportional to the local number density of
rods, while the tensor $\mathbf{A}_{2}$ describes the development
of orientational order during demixing. As will turn out, the
number density $A_{0}$ is enslaved by the orientational
contribution $\mathbf{A}_{2}$.

Using these definitions in the equation of motion Eq.(\ref{P1}),
the corresponding equations of motion for $A_{0}$ and
$\mathbf{A}_{2}$ can be derived. These equations can be solved,
leading to,
\begin{eqnarray} \label{final}
A_{0}(\kk,t)&=&-\, \frac{
\mbox{\small{$\frac{1}{10}$}}\,\bar{D}\,k^{2}\,\left\{1
-\mbox{\small{$\frac{1}{4}$}}\,
\mbox{\small{$\frac{L}{D}$}}\,\varphi\left(\,1+\mbox{\small{$\frac{29}{84}$}}\,(kL)^{2}\right)\right\}}{
\bar{D}\,k^{2}\,\left\{\,1 +
2\mbox{\small{$\frac{L}{D}$}}\,\varphi \,\right\}- 6D_{r}\left\{
\,
1-\mbox{\small{$\frac{1}{4}$}}\,\mbox{\small{$\frac{L}{D}$}}\,\varphi\,\left(
1-\mbox{\small{$\frac{499}{8064}$}}\,(kL)^{2}\right)\right\}}\;
\hkk\hkk\!:\!\mathbf{A}_{2}(\kk,t)\;,\nonumber \\
\hkk\hkk\!:\!\mathbf{A}_{2}(\kk,t)&=&\hkk\hkk\!:\!\mathbf{A}_{2}(\kk,t=0)\,
\exp\left\{- \lambda^{(-)}\,t\right\}\;,
\end{eqnarray}
where $\varphi$ is the volume fraction of rods, and $L$ and $D$
are their length and thickness, respectively. Furthermore, $\hkk$
is the unit vector along the wave vector $k$, and $\lambda^{(-)}$
is the eigen value related to the unstable mode, which is equal
to,
\begin{eqnarray} \label{without3}
\lambda^{(-)}&=&6D_{r}\left\{ \,
1-\mbox{\small{$\frac{1}{4}$}}\,\mbox{\small{$\frac{L}{D}$}}\,\varphi\,\left(
1-\mbox{\small{$\frac{499}{8064}$}}\,(kL)^{2}\right)\right\}-\mathcal{O}(kL^{4})\;.
\end{eqnarray}

Note that for $\mbox{\small{$\frac{L}{D}$}}\,\varphi>4$, this
eigen value is negative for sufficiently small wave vectors, so
that, according to eq.(\ref{final}), inhomogeneities with the
corresponding wave length $2\pi/k$ will grow in time without any
time delay. This concentration marks the location of the
isotropic-to-nematic spinodal and is in accordance with Onsager's
prediction \cite{Onsager49}.

The proportionality of the density $A_{0}$ to the orientational
contribution $\mathbf{A}_{2}$ in Eq.(\ref{final}) reflects the
enslavement of density to orientational order during demixing.
That is, the transition is driven by orientational diffusion
rather than translational diffusion.

Although Eq.(\ref{without3}) has been derived for an initial
isotropic distribution, we think that the main conclusion, i.e.
orientational fluctuations dominate the phase separation, is also
valid for phase separation starting from the nematic state.

In a scattering experiment the total scattered intensity is
related to the quantities $A_{0}$ and
$\hkk\hkk\!:\!\mathbf{A}_{2}$ as,
\begin{eqnarray} \label{scatter2}
I\;\sim\;\left[\,A_{0}(\kk,t)\,\left(1-\mbox{\small{$\frac{1}{72}$}}\,(kL)^{2}\right)-
\mbox{\small{$\frac{1}{180}$}}\,(kL)^{2}\,\hkk\hkk:\mathbf{A}_{2}(\kk,t)\,\left(\,1-
\mbox{\small{$\frac{3}{560}$}}\,(kL)^{2}\,\right)+\mathcal{O}\left((kL)^{6}\right)\,\right]^{2}\;.
\end{eqnarray}
As for gas-liquid spinodal demixing suspensions of spheres, the
scattered intensity during isotropic-nematic demixing suspensions
of rods exhibits a ring-like pattern where a particular finite
wavevevector grows most rapidly. The occurrence of a maximum in
the scattered intensity at finite wavevectors during demixing has
a fundamentally different origin for demixing rods as compared to
spheres. For gas-liquid demixing of suspensions of spheres, the
eigenvalue (here referred to as $\lambda^{(-)}$) itself exhibits
an extremum at a finite wavevector. For spheres $\lambda^{(-)}$ is
of the form $D\,k^{2}\,[ 1-\alpha\,k^{2}]$, where $D$ and $\alpha$
are wavevector-independent, positive coefficients
\cite{Cahn65,Kojima99}. The prefactor $k^{2}$ signifies the fact
that diffusion of spheres over long distances takes longer times,
while $\alpha\,k^{2}$ signifies the stabilization of large
concentrations gradients. It is easily verified that an eigenvalue
of this form exhibits an extremum at a finite wavevector. For
gas-liquid demixing of suspensions of spheres this maximum in the
growth rate results in the maximum in the scattering pattern. For
the isotropic-nematic demixing of suspensions of \emph{rods}, the
eigenvalue is of the form $D\,[ 1-\alpha\,k^{2}]$ (see
eq.(\ref{without3}), that is, the prefactor $k^{2}$ as compared to
spheres is missing here. As a consequence $\lambda^{(-)}$ remains
finite at zero wavevectors. This difference in wavevector
dependence of the eigenvalue $\lambda^{(-)}$ for gas-liquid and
isotropic-nematic demixing is due to the fact that gas-liquid
demixing is governed by translational diffusion while
isotropic-nematic demixing is (predominantly) governed by
rotational diffusion. The maximum in the scattering pattern is now
due to the combination of the wavevector dependence of the
eigenvalue {\it and} the wavevector dependent prefactors to the
time-exponent. Note that according to eq.(\ref{scatter2}), the
prefactor of the time-exponent is indeed $\sim\,k^{2}$, rendering
\emph{the scattered intensity equal to zero at zero wavevector},
which expresses conservation of the number of rods. The
$k$-dependence of $\lambda^{(-)}$ can be tested experimentally
using the fact that, according to
Eqs.(\ref{final},\ref{scatter2}), the scattered intensity is $\sim
\exp\{\lambda^{(-)}t\}$. Hence,
\begin{eqnarray} \label{experiment1}
\frac{\partial\;}{\partial
t}\;\ln\{I(k,t)\}\;=\;2\,\lambda^{(-)}\;=\;12\,D_{r}\left\{ \,
1-\mbox{\small{$\frac{1}{4}$}}\,\mbox{\small{$\frac{L}{D}$}}\,\varphi\,\left(
1-\mbox{\small{$\frac{499}{8064}$}}\,(kL)^{2}\right)\right\}\;.
\end{eqnarray}
The slope of a plot of $\ln\{I(k,t)\}$ as a function of $t$ for a
given wave vector is thus equal to $\lambda^{(-)}$ for that
particular wave vector. Repeating this for various wave vectors
allows to construct the wave vector dependence of $\lambda^{(-)}$.

In addition, the critical wave vector $k_{c}$ above which the
system becomes stable, that is, where $\lambda^{(-)}$ becomes
positive, is equal to,
\begin{eqnarray} \label{critical}
k_{c}\,L\;=\;2\,\sqrt{\frac{8064}{499}}\;
\sqrt{\frac{1}{4}-\frac{1}{\mbox{\small{$\frac{L}{D}$}}\,\varphi}}\;\approx\;
8\,\sqrt{\frac{1}{4}-\frac{1}{\mbox{\small{$\frac{L}{D}$}}\,\varphi}}\;.
\end{eqnarray}
For shallow quenches, that is, for concentrations where the
$\mbox{\small{$\frac{L}{D}$}}\,\varphi$ is close to $4$, the
critical wave vector is thus relatively small. That is, \emph{
shallow quenches result in relatively large scale inhomogeneities
while deeper quenches give rise to relatively small scale
inhomogeneities.}

In this paper we prepare an initial nematic state, by shearing a
suspension at large enough shear rate such that the induced
nematic phase is stable against phase separation (see
ref.\cite{Dhont03c} for a discussion of the bifurcation diagram
for sheared systems), and then quench to zero shear rate. Since
the orientation of the rods dominates the phase separation, it is
expected that \emph{phase separation takes place anisotropically
for an initial nematic state}. In the following we will test this
assumption and also the predictions made above for the
isotropic-nematic SD, which we believe to hold true also for the
nematic-isotropic SD.

\section{Materials and Methods}\label{Materials and methods}

As model colloidal rods we use \emph{fd}-virus particles which
were grown as described in ref.\cite{Dogic04a}. The physical
characteristics of the bacteriophage \emph{fd} are: length $L=
880\; nm$; diameter $D= 6.6\; nm$; persistence length $2.2 \;\mu
m$. A homogeneous solution of $22.0 \;  mg/mL$ \emph{fd}-virus and
$10.6 \; mg/mL$ of dextran (507 kd, Sigma-Aldrich, radius of
gyration of $18\;nm$) in $20 mM$ tris buffer at $pH\;8.15$ with
$100 mM$ NaCl is allowed to macroscopically phase separate.
Without dextran the binodal concentrations are $21 $ and $23 \;
mg/mL$ for the isotropic and nematic phase, respectively. Due to
the added dextran, the binodal points shift to $17$ and $31\;
mg/mL$, respectively. The lower spinodal point $C^{spin}$ for this
sample was found to be equal to 24.7$\pm$1.1 $mg/mL$, as
determined in a previous paper \cite{Lettinga05c}. We prepared
three dispersions by mixing a known volume of coexisting isotropic
and nematic bulk phases of the quiescent dispersion. In this way,
the osmotic pressure is independent of the varying ratio of
dextran to \emph{fd}-virus concentration. The mixing ratios of
isotropic and nematic phases are chosen such that a quench from
the nematic phase under flow will always render the aligned system
unstable without flow, that is, the \emph{fd} concentration is
larger than the lower binodal concentration $C^{bin}=17\;mg/mL$
and smaller than the lower spinodal concentration
$C^{spin}=24.7\;\pm 1.1\;mg/mL$ \cite{Lettinga05c}. The \emph{fd}
concentrations are denoted here after as $\phi_{f}$, where
$f=(C-C^{bin})/(C^{spin}-C^{bin})$ relates to the fraction of the
concentration between the lower binodal and spinodal. The
concentrations used in the present study are : $\phi_{0.52} = 19.3
\;mg/mL$, $\phi_{0.55} = 19.9 \;mg/mL$ and $\phi_{0.84} = 23.6
\;mg/mL$. For the SALS measurements the concentration of \emph{fd}
and dextran were $21.0 \; mg/mL$ and $12.1\; mg/mL$, respectively.
Due to the fact that we used different concentrations of dextran
for the different experiments, we cannot directly compare SALS
data to the microscopy data.

For the microscopy experiments we used a home-built counter
rotating cone-plate shear cell, placed on top of a Leica TCS-SP2
inverted confocal microscope. This cell has a plane of zero
velocity in which objects remain stationary with respect to the
microscope while shearing. For details of the setup we refer to
ref.\cite{Derks04}. For the measurements described here we used
confocal reflection mode at a wavelength of $488\, nm$. Quench
experiments were done as follows. Samples were first sheared at a
high rate of $10\; s^{-1}$ for several minutes. The shear was then
suddenly stopped, after which images were recorded at regular time
intervals. For the SALS measurements we used a home made
cylindrical optical shear cell. The rotating hollow inner cylinder
has a radius of $21.5\, mm$, the gap width is $2.47\, mm$. The
shear cell is placed in a cylindrical toluene bath with the second
gap of the cell exactly in the middle of the bath. A $5\, mW$
He-Ne laser (Melles-Griott) with a wavelength of $632.8\, nm$ was
used as a light source. The laser beam is directed along the
gradient direction through one single gap using a periscope
system, which is inserted into a silicon oil filled inner cylinder
at a fixed position. In this way the flow-vorticity plane is
probed. Scattered intensities are projected on a white screen,
with a beam-stop in the middle . The size of the beam-stop
corresponds to a scattering angle of $1.4^{o}$ and a wave vector
of $2.4 \times 10^{5}\, m^{-1}$. Images were taken in transmission
with a peltier cooled 12 bit CCD camera, with 582x782 pixels
(Princeton Instruments, microMAX). The maximum scattering angle
was $5.1^{o}$, which corresponds to a wave vector of $1.0\times
10^{6}\, m^{-1}$.

\begin{figure}
\includegraphics[width=.85\textwidth]{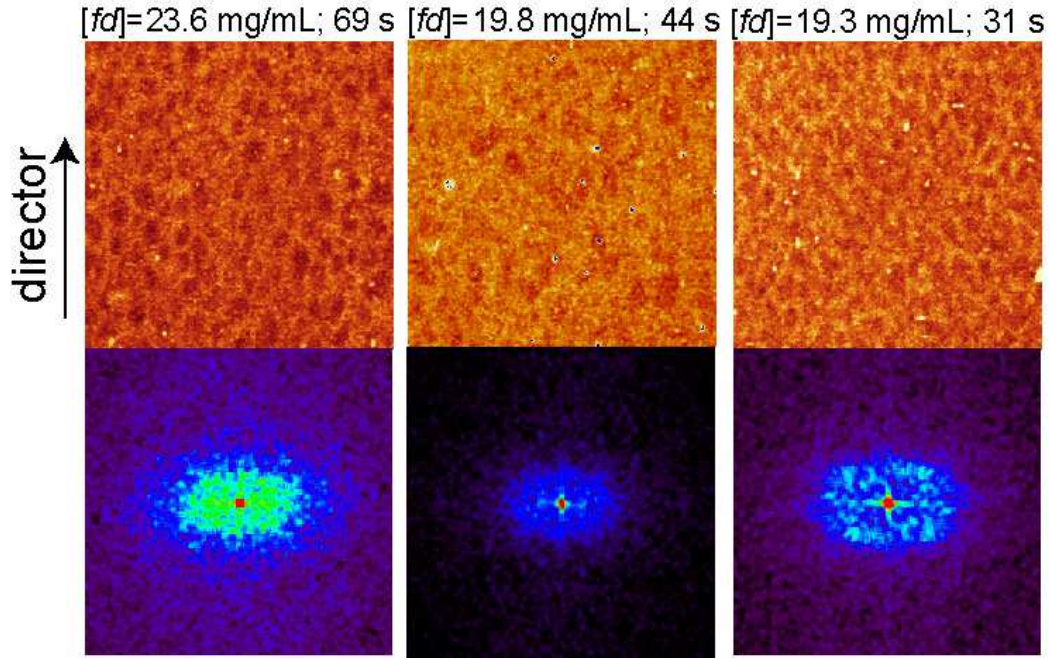}
\caption{\label{initial} (Color) The initial stages of SD for
three different concentrations, as indicated in the figure. The
top row shows the micrographs taken by reflection confocal
scanning laser microscopy (field of view $=110\;\mu m$); the
bottom row shows the fourier transform of the micrographs.}
\end{figure}

\begin{figure}
\includegraphics[width=.55\textwidth]{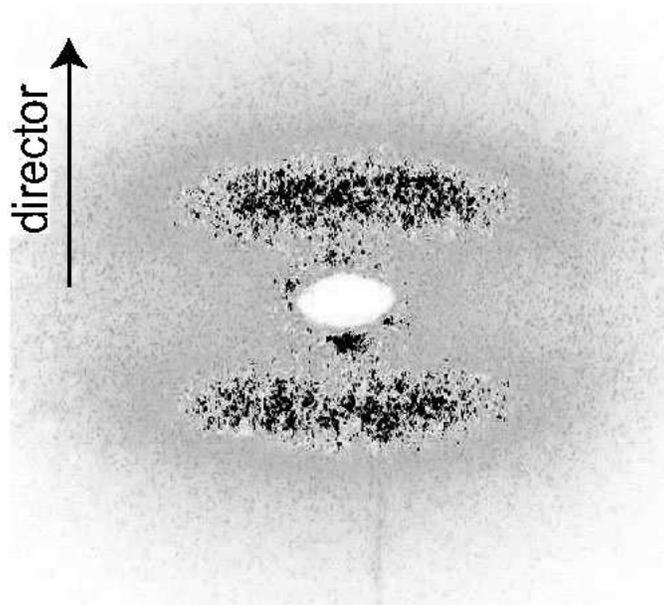}
\caption{\label{SALS} The scattering pattern for a sample with
[\emph{fd}]$=21.0 \; mg/mL$ and [dextran]$=12.1 \; mg/mL$ taken $
23 \; s$ after cessation of flow.}
\end{figure}

\section{Experimental Results}

Confocal images of the morphology during spinodal decomposition in
the early stage after a shear rate quench for different
concentrations are given in the top row in Fig.\ref{initial} while
in the bottom the corresponding Fourier transforms are plotted.
The observed ring-like scattering patterns, typical for spinodal
decomposition, are anisotropic with symmetry around the nematic
director which is along the flow direction. The same anisotropy is
observed in the SALS measurements, where such Fourier space images
are directly probed (see Fig.\ref{SALS}).

\begin{figure}
\includegraphics[width=.55\textwidth]{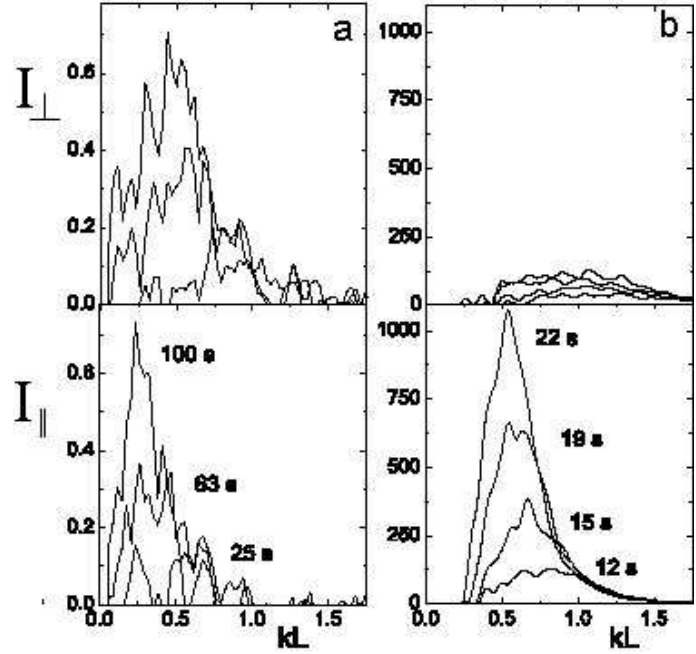}
\caption{\label{FFT_profile} Cross sections of the Fourier
transform of the confocal images parallel and perpendicular to the
director for a \emph{fd}-concentration of $\phi_{0.52}$ (a), and
the parallel and perpendicular cross sections of the scattered
intensity as found from SALS (b). The wave vector is scaled by the
length $L$ of the \emph{fd}-virus, the intensities are given in
arbitrary units.}
\end{figure}

\begin{figure}
\includegraphics[width=.95\textwidth]{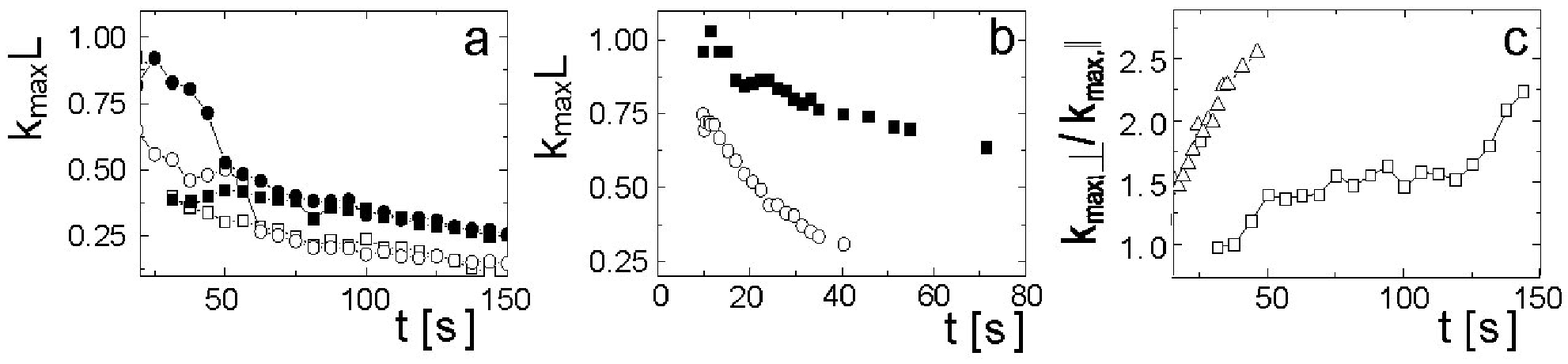}
\caption{\label{qmax_qratio} The wavelength of the fastest growing
Fourier component $k_{max}L$ as found from the cross sections
parallel to the director (open symbols) and perpendicular to the
director (filled symbols) of (a) the Fourier transforms of
confocal images with a \emph{fd}-concentration of $\phi_{0.84}$
(squares) and $\phi_{0.52}$ (circles), and (b) the SALS pattern.
(c) The anisotropy of the spinodal structure given by the ratio
$k_{max,\bot} L /k_{max,\|} L$ for $\phi_{0.84}$ (squares)
 and the SALS measurement (triangles).}
\end{figure}

Cross sections of the Fourier transforms and scattering patterns
parallel and perpendicular to the director are given in
Fig.\ref{FFT_profile} a and b, respectively. From these profiles
we obtain the wave vector at which the Fourier transform exhibits
its maximum, $k_{max}L$, quantifying the wavelength of the fastest
growing Fourier component of the inhomogeneous morphology. The
values for $k_{max}L$ for directions parallel and perpendicular to
the director as obtained from such cross sections of the Fourier
transform of the confocal images and the SALS patterns are plotted
as a function of time in Fig.\ref{qmax_qratio}a and b,
respectively. The ratio $k_{max,\,\perp}/k_{max,\,\parallel}$ of
$k_{max}$ perpendicular and parallel to the director is plotted in
Fig.\ref{qmax_qratio}c. This figure thus characterizes the
anisotropy in the morphology. Since this ratio is larger than one,
the typical size of inhomogeneities perpendicular to the director
is smaller than the size of inhomogeneities in the direction
parallel to the director.

Comparing the profiles obtained from microscopy and from SALS, it
is obvious that the SALS signal is less noisy. This is due to the
fact that the volume that is being probed by microscopy is much
smaller than the volume probed by SALS. On the other hand, due to
the beam stop the profile starts at higher k-values than the
profiles obtained from microscopy. The large difference in the
scattered intensity parallel and perpendicular to the flow
direction is also observed under identical conditions for
isotropic systems, and is caused by the experimental
set-up\cite{Holmqvist05}.

\section{Discussion}

The main result of the theoretical treatment in section
\ref{theory} for SD of a dispersion of rods is expressed by
Eq.(\ref{final}). This equation shows that changes in the
concentration of rods are enslaved by changes in the orientation
of the rods. Although the treatment was done for an initially
isotropic state we believe that this result is valid independent
of the initial condition. This relation between concentration and
orientation has a few consequences, which can be tested
experimentally:

1) as for spheres there is no demixing for $k\rightarrow 0$;

2) but the driving force of phase separation given by
$\lambda^{(-)}$ goes to a constant value for $k\rightarrow 0$ and
not to zero as would have been the case when translational
diffusion dominates phase separation kinetics (see the exponent in
Eq.(\ref{final})). As a result $k_{max}$ shifts in time also in
the initial stage of demixing;

3) shallow quenches result in relatively large scale
inhomogeneities while deeper quenches give rise to relatively
small scale inhomogeneities, see Eq.(\ref{critical}).

All micrographs in images Fig.\ref{initial} show interconnected
structures typical for spinodal decomposition. The resulting
Fourier transforms in Fig.\ref{initial} as well as the SALS
pattern in Fig.\ref{SALS} are slightly elongated in the direction
of the director. The observation of a ring structure confirms the
prediction that the scattered intensity is zero for $k\rightarrow
0$, i.e. that there is no demixing for $k\rightarrow 0$ (see also
the cross sections in Fig.\ref{FFT_profile}). As pointed out in
the theory section this is a general feature of SD, and it is a
consequence of the conservation of the number of rods. The
anisotropy in the morphology as well as in the growth rates, see
Fig.\ref{qmax_qratio}c, show that the formation of inhomogeneities
is affected by the initial orientation of the rods. The anisotropy
in the phase separation is also seen in the case of
nucleation-and-growth, where we observed that the nucleating
tactoids of isotropic phase are oriented along the director of the
nematic background phase \cite{Lettinga05c}. This anisotropy is
due to residual alignment after the quench of the initially
strongly sheared suspension.

\begin{figure}
\includegraphics[width=.95\textwidth]{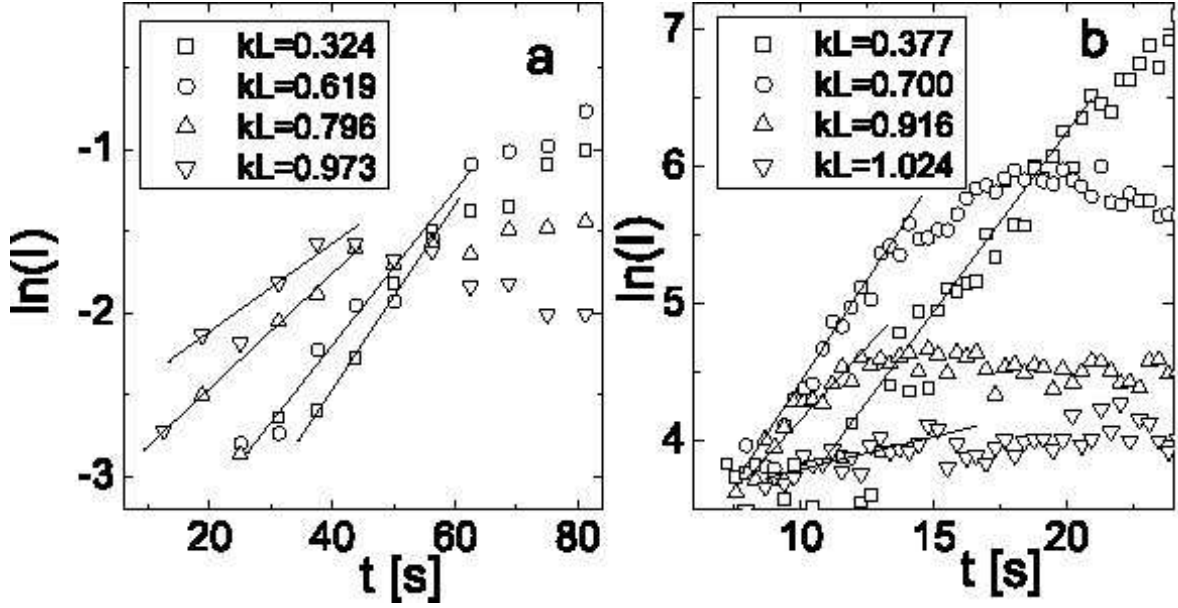}
\caption{\label{lnIfft} The logarithm of the Fourier component of
confocal images for sample $\phi_{0.52}$ (a) and the scattered
intensity from SALS experiments (b) as a function of time for
various values of $kL$. The eigen value $\lambda^{(-)}$ is
obtained from the initial slope of this plot.}
\end{figure}

When plotting the wave vector where the intensity ring exhibits
its maximum, i.e. $k_{max}$, as a function of time, it is readily
seen that this maximum shifts to smaller values also during the
initial stage of demixing right after the quench (see Fig.
\ref{qmax_qratio}). That the initial stage of demixing is probed
follows from Fig.\ref{lnIfft}, where the logarithm of the
intensity is plotted versus time. In the initial stage this
relation should be linear (see Eq.(\ref{experiment1})), which is
indeed seen to be the case for all values of $kL$. The initial
stage of the SD ends where this curve starts to deviate from
linearity. For more shallow quenches closer to the spinodal, at
higher concentrations, the initial stage extends up to 100
seconds.

\begin{figure}
\includegraphics[width=.95\textwidth]{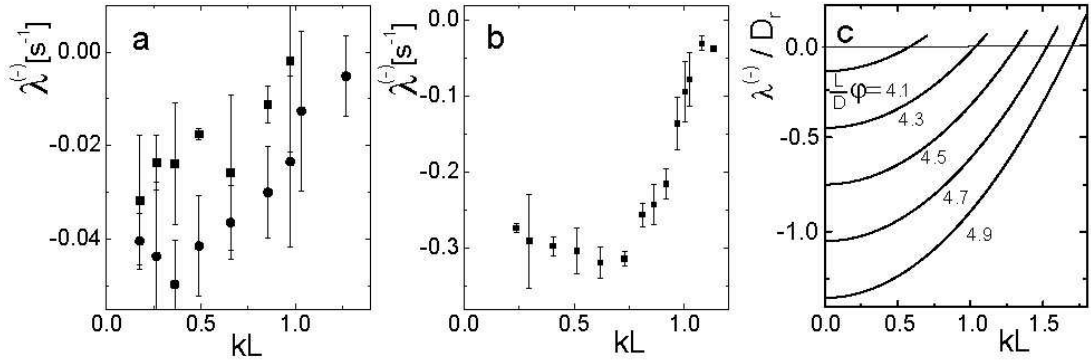}
\caption{\label{lambdas} The $kL$-dependence of the eigen value
$\lambda^{(-)}$ as obtained from confocal microscopy for
$\phi_{0.52}$ (bullets) and $\phi_{0.54}$ (squares) (a),  as
obtained from SALS (b), and as obtained from theory for the
isotropic-nematic transition for different dimensionless
concentrations $\frac{L}{D}\phi$(c). Here $L$ and $D$ are the rod
length and thickness, respectively, $\phi$ is the volume fraction
of the rods, and $D_r$ is its rotational diffusion at infinite
dilution.}
\end{figure}

More importantly, for each value of  $kL$ we obtain the phase
separation rate $\lambda^{(-)}$ from the slope of the curve (see
Eq.(\ref{experiment1})). The resulting curves of $\lambda^{(-)}$
vs. $kL$ are plotted in Fig.\ref{lambdas} for two different
confocal microscopy samples (a) and the SALS sample (b). Clearly,
$\lambda^{(-)}$ approaches a finite value for $k\rightarrow 0$, as
was predicted by theory, see Fig.\ref{lambdas}c. This shows that
the demixing kinetics is dominated by rotational diffusion. The
absolute value for $\lambda^{(-)}$ is about a factor ten higher
for the SALS experiment than for the microscopy experiments, which
is probably due to the lower dextran concentration that has been
used in the microscopy experiment as compared to the SALS
experiment. The analysis we used for our data could in principle
also be applied to the measurements of van Bruggen et al.
\cite{vanbruggen99a}, where a similar behavior of the SALS
patterns is observed.

Fig.\ref{qmax_qratio}a also confirms the theoretical prediction in
Eq.(\ref{critical}) that the initial inhomogeneities are larger
for shallow quenches than for deep quenches. The length scale of
the initially formed structures for the deep quench
(circles,$\phi_{0.84}$) is 11 times the rod length in the
direction parallel to the initial director and 7.1 times the rod
length in the direction perpendicular to the initial director. For
the shallow quench (squares, $\phi_{0.52}$) the initial structure
is barely anisotropic and has a typical size of 17 times the rod
length. After about one minute the typical sizes for both
concentrations start to overlap.

Finally we would like to remind that the theory presented in
section \ref{theory} is valid for repulsive rods. In our
experimental system, however, depletion attractions between rods
are induced by adding dextran. On adding more dextran,
translational diffusion could play a more important role during
phase separation. Rod-polymer mixtures has been treated on the
basis of Ginzburg-Landau equations of motion with a thermodynamics
input, amongst others, by Liu and Fredrickson \cite{Liu96},
Matsuyama et al.\cite{Matsuyama00}, and by Fukuda \cite{Fukuda99}.
In these papers it is shown that translational diffusion indeed
becomes more important on increasing the polymer concentration.
This results in a minimum in $\lambda^{(-)}$ as a function of the
wave vector at sufficiently high polymer concentration. We
therefore believe that the pronounced minimum that is observed for
the SALS measurement, see Fig.\ref{lambdas}b, is related to
attractions between the rods as induced by the added dextran. The
two microscopy measurements depicted in  Fig.\ref{lambdas}a are
done at a lower overall dextran concentration. The sample with the
somewhat lower \emph{fd} concentration, i.e. higher polymer
concentration, shows a less pronounced minimum as compared to the
SALS sample, while for the sample with the lowest polymer
concentration no minimum is present. These observations confirm
the theoretical prediction on the polymer dependence of the phase
separation kinetics. In future studies we will systematically vary
the polymer concentration for a fixed concentration of \emph{fd}.

\section{Conclusion}

We studied the nematic-isotropic SD of dispersions of
\emph{fd}-virus particles with added polymer after shear quenches
into the two-phase region for varying concentrations. We
exemplified the fundamental difference between spinodal
decomposition of dispersions of rods and spheres using a recently
developed theory. The main difference is that in the case of rods
the phase separation is dominated by rotational diffusion. As a
result the eigen value  $\lambda^{(-)}$, which quantifies the rate
of the phase separation, approaches a non-zero constant value for
$k \rightarrow 0$, contrary to gas-liquid demixing of spheres
where the corresponding eigen value becomes zero for $k
\rightarrow 0$. This is due to the fact that for rods a local
reorientation is sufficient to start the phase separation, whereas
for spheres translational diffusion over finite distances is
needed. We found experimentally the same $k$-dependence of
$\lambda^{(-)}$ as predicted by theory
\cite{Dhont05,Liu96,Matsuyama00}. Our experiments thus confirm
that demixing is dominated by rotational diffusion and not by
translational diffusion as suggested in earlier work
\cite{winters00,maeda89}. In addition, we found a possible effect
of translational diffusion through the minimum of the wave vector
dependence of the unstable eigen mode, due to attractions between
the rods as induced by the added dextran refs.
\cite{Liu96,Matsuyama00}. This will be subject of further
investigations.

\section*{Acknowledgement}
This work was performed within the framework of the Transregio SFB
TR6 "Physics of colloidal dispersions in external fields" and the
European Network of Excellence Soft Matter Composites (SoftComp).


\begin{thebibliography}{10}

\bibitem{Onsager49}
L.~Onsager.
\newblock The effect of shape on the interaction of colloidal particles.
\newblock {\em Annals of the New York academy of science}, 51:62--659, 1949.

\bibitem{Vroege92}
G.~J. Vroege and H.~N.~W. Lekkerkerker.
\newblock Phase transitions in lyotropic colloidal and polymer liquid crystals.
\newblock {\em Rep. Prog. Phys.}, 55:1241--1309, 1992.

\bibitem{Bolhuis97}
P.~Bolhuis and D.~Frenkel.
\newblock Tracing the phase boundaries of hard spherocylinders.
\newblock {\em J. Chem. Phys.}, 106(2):666--687, 1997.

\bibitem{Graf99}
H.~Graf and H.~L\"{o}wen.
\newblock Phase diagram of tobacco mosaic virus solutions.
\newblock {\em Phys. Rev. E}, 59(2):1932--1942, 1999.

\bibitem{Kayser78}
R.~F.~Kayser Jr. and H.~J. Ravech\'{e}.
\newblock Bifurcation in onsager's model of the isotropic-nematic transition.
\newblock {\em Phys. Rev. A}, 17:2067--2072, 1978.

\bibitem{Dhont05}
J.~K.~G. Dhont and W.~J. Briels.
\newblock Isotropic-nematic spinodal decomposition kinetics.
\newblock {\em Phys. Rev. E}, 72:031404, 2005.

\bibitem{winters00}
J.~W. Winters, Th. Odijk, and P.~van~der Schoot.
\newblock Spinodal decomposition in a semidilute suspension of rodlike
  macromolecules.
\newblock {\em Phys. Rev. E}, 63:011501--1, 2000.

\bibitem{maeda89}
T.~Maeda.
\newblock Matrix representation of the dynamical structure factor of a solution
  of rodlike polymers in the isotropic phase.
\newblock {\em Macromolecules}, 22:1881--1891, 1998.

\bibitem{Liu96}
A.~J. Liu and G.~H. Fredrickson.
\newblock Phase separation kinetics of rod/coil mixtures.
\newblock {\em Macromolecules}, 29:8000--8009, 1996.

\bibitem{Matsuyama00}
A.~Matsuyama, R.~M.~L. Evans, and M.~E. Cates.
\newblock Orientation fluctuation-induced spinodal decomposition in
  polymer–liquid-crystal mixtures.
\newblock {\em Phys. Rev. E}, 61(3):2977--2986, 2000.

\bibitem{vanbruggen99a}
M.~P.~B. van Bruggen, J.~K.~G. Dhont, and H.~N.~W. Lekkerkerker.
\newblock Morphology and kinetics of the isotropic-nematic phase transition in
  dispersions of hard rods.
\newblock {\em Macromolecules}, 32:2256--2264, 1999.

\bibitem{Viamontes05}
J.~Viamontes and J.~X. Tang.
\newblock Formation of nematic liquid crystalline phase of f-actin varies from
  continuous to biphasic transition.
\newblock {\em http://arxiv.org/abs/cond-mat/0506813}, 2005.

\bibitem{Lenstra01}
T.~A.~J. Lenstra, Z.~Dogic, and J.~K.~G. Dhont.
\newblock Shear-induced displacement of isotropic-nematic spinodals.
\newblock {\em J. Chem. Phys.}, 114(22):10151--10162, 2001.

\bibitem{Tang93}
J.~Tang and S.~Fraden.
\newblock Magnetic-field-induced phase transition in a colloidal suspension.
\newblock {\em Phys. Rev. Lett.}, 71(21):3509--3512, 1993.

\bibitem{Tang95}
J.~Tang and S.~Fraden.
\newblock Isotropic-cholesteric phase transition in colloidal dispersions of
  filamentous bacteriophage \textit{fd}.
\newblock {\em Liquid Crystals}, 19(4):459--467, 1995.

\bibitem{Chen93}
Z.~Y. Chen.
\newblock Nematic ordering in semiflexible polymer chains.
\newblock {\em Macromolecules}, 26:3419--3423, 1993.

\bibitem{Dogic04a}
Z.~Dogic, K.~R. Purdy, E.~Grelet, M.~Adams, and S.~Fraden.
\newblock Isotropic-nematic phase transition in suspensions of filamentous
  virus and dextran.
\newblock {\em Phys. Rev. E}, 69:051702, 2004.

\bibitem{Lettinga05c}
M.P. Lettinga, K.~Kang, A.~Imhof, D.~Derks, and J.K.G. Dhont.
\newblock Kinetic pathways of the nematic-isotropic phase transition of
  rod-like viruses.
\newblock {\em J. Phys.: Condens. Matter}, 17:S3609–S3618, 2005.
\newblock {The dextran concentration given in this paper was for hydrated dextran.
Dehydrated dextran had a concentration of $10.6 \; mg/mL$. }

\bibitem{Doi86}
M.~Doi and S.~F. Edwards.
\newblock {\em The Theory of Polymer Dynamics}.
\newblock Oxford, 1986.

\bibitem{Cahn65}
J.~W. Cahn.
\newblock Phase separation by spinodal decomposition in isotropic systems.
\newblock {\em J. Chem. Phys.}, 42(1):93--99, 1965.

\bibitem{Kojima99}
J.~Kojima, M.~Takenaka, Y.~Nakayama, and T.~Hashimoto.
\newblock Early stage spinodal decomposition in polymer solution under high
  pressure.
\newblock {\em Macromolecules}, 32:1809--1815, 1999.

\bibitem{Dhont03c}
J.~K.~G. Dhont and W.~J. Briels.
\newblock Viscoelasticity of suspensions of long, rigid rods.
\newblock {\em Colloid Surface A}, 213(2-3):131--156, 2003.

\bibitem{Derks04}
D.~Derks, H.~Wisman, A.~van Blaaderen, and A.~Imhof.
\newblock Confocal microscopy of colloidal dispersions in shear flow using a
  counter-rotating cone–plate shear cell.
\newblock {\em J. Phys.: Condens. Matter}, 16:S3917--S3927, 2004.

\bibitem{Holmqvist05}
P.~Holmqvist, M.~P. Lettinga, J.~Buitenhuis, and J.~K. G. Dhont.
\newblock Crystallization Kinetics of Colloidal Spheres under
Stationary Shear Flow.
\newblock {\em Langmuir}, 21:10976-10982, 2005.

\bibitem{Fukuda99}
J.~Fukuda.
\newblock Phase separation kinetics of liquid crystalline polymers: Effect of
  orientational order.
\newblock {\em Phys. Rev. E}, 59(3):3275--3288, 1999.

\end{thebibliography}
\end{document}